\pdfoutput=1
\documentclass[10pt]{article}
\usepackage{geometry}
\usepackage{xcolor}
\definecolor{graycolor}{gray}{0.9}
\usepackage{microtype}
\usepackage{setspace}
\usepackage{cmap}
\usepackage[utf8]{inputenc}
\usepackage[T1]{fontenc}
\usepackage[english]{babel}
\usepackage{times}
\usepackage{array,amsthm,amssymb,amsmath}
\def\K{{\cal K}}

\makeatletter
\renewcommand{\maketag@@@}[1]{\hbox{\m@th\normalsize\normalfont#1}}
\makeatother
\usepackage{soul}
\usepackage{cite}
\usepackage[numbers,sort&compress]{natbib}

\usepackage{titlesec}
\titleformat {\section} [block] {\raggedright \fontsize{10}{10}\selectfont\bfseries} {\thesection. \space} {0pt} {}
\titlespacing {\section} {0pt} {12pt} {6pt}
\titleformat {\subsection} [block] {\raggedright \fontsize{10}{10}\selectfont\itshape} {\thesubsection .\space} {0pt} {}
\titlespacing {\subsection} {0pt} {12pt} {6pt}
\titleformat {\subsubsection} [block] {\raggedright \fontsize{10}{10}\selectfont} {\thesubsubsection .\space} {0pt} {}
\titlespacing {\subsubsection} {0pt} {12pt} {6pt}
\titleformat {\paragraph} [block] {\raggedright \fontsize{10}{10}\selectfont} {} {0pt} {}
\titlespacing {\paragraph} {0pt} {12pt} {6pt}

\usepackage{array} \newcommand{\PreserveBackslash}[1]{\let\temp=\\#1\let\\=\temp}
\newcolumntype{C}[1]{>{\PreserveBackslash\centering}m{#1}}
\newcolumntype{R}[1]{>{\PreserveBackslash\raggedleft}m{#1}}
\newcolumntype{L}[1]{>{\PreserveBackslash\raggedright}m{#1}}
\usepackage{lineno}
\usepackage{tabularx}
\usepackage{colortbl}
\usepackage{graphicx}
\usepackage{float}
\usepackage[export]{adjustbox}
\usepackage{caption}
\usepackage{fancyhdr}
\usepackage{lastpage}
\usepackage{layout}
\usepackage{setspace}
\usepackage{enumitem}
\usepackage{booktabs}
\usepackage{arydshln}
\usepackage{multirow}
\usepackage{color}
\usepackage{hyperref}
\hypersetup{
	colorlinks=true,
	linkcolor=blue,
	filecolor=blue,
	urlcolor=black,
	citecolor=cyan,
}

\fancypagestyle{firstpage}{
    \setlength{\headsep}{2.2cm}
    
    \setlength{\footskip}{1.5cm}
    \fancyhf{}% clear default for head and foot
    \lhead{\begin{table}[H]
        \centering
        \begin{tabular}{L{2.5cm}C{10cm}C{3.1cm}R{2cm}}
            \includegraphics[scale=0.035]{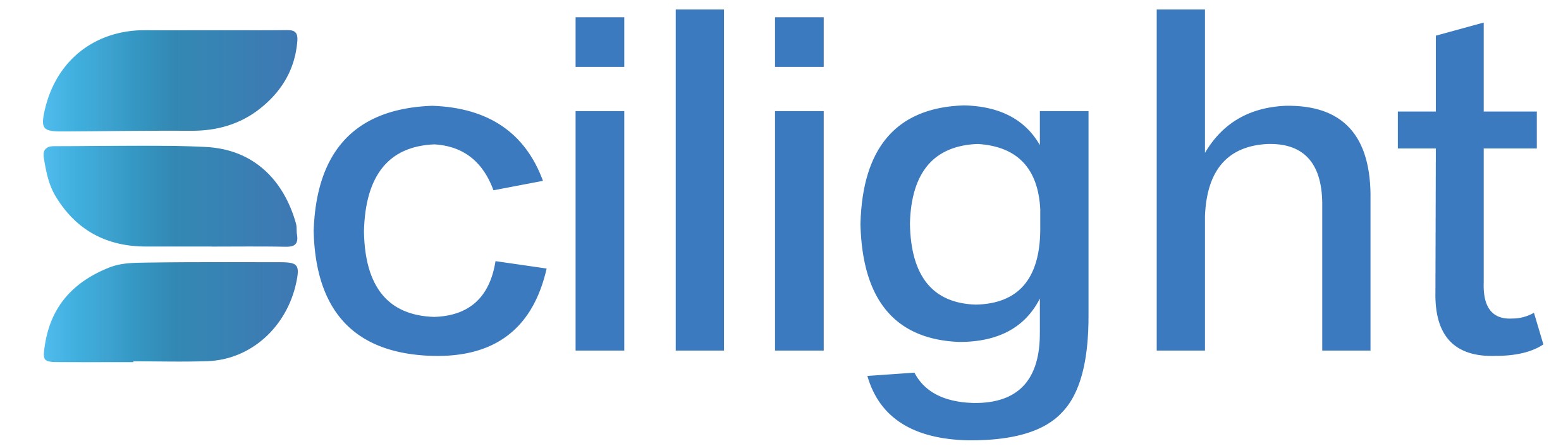} \vspace{-6pt}& \cellcolor{graycolor}\begin{tabular}[c]{@{}c@{}}\textit{International Journal of Gravitation and Theoretical Physics}\\ \href{https://www.sciltp.com/journals/ijgtp}{https://www.sciltp.com/journals/ijgtp}\end{tabular} & \includegraphics[scale=0.020]{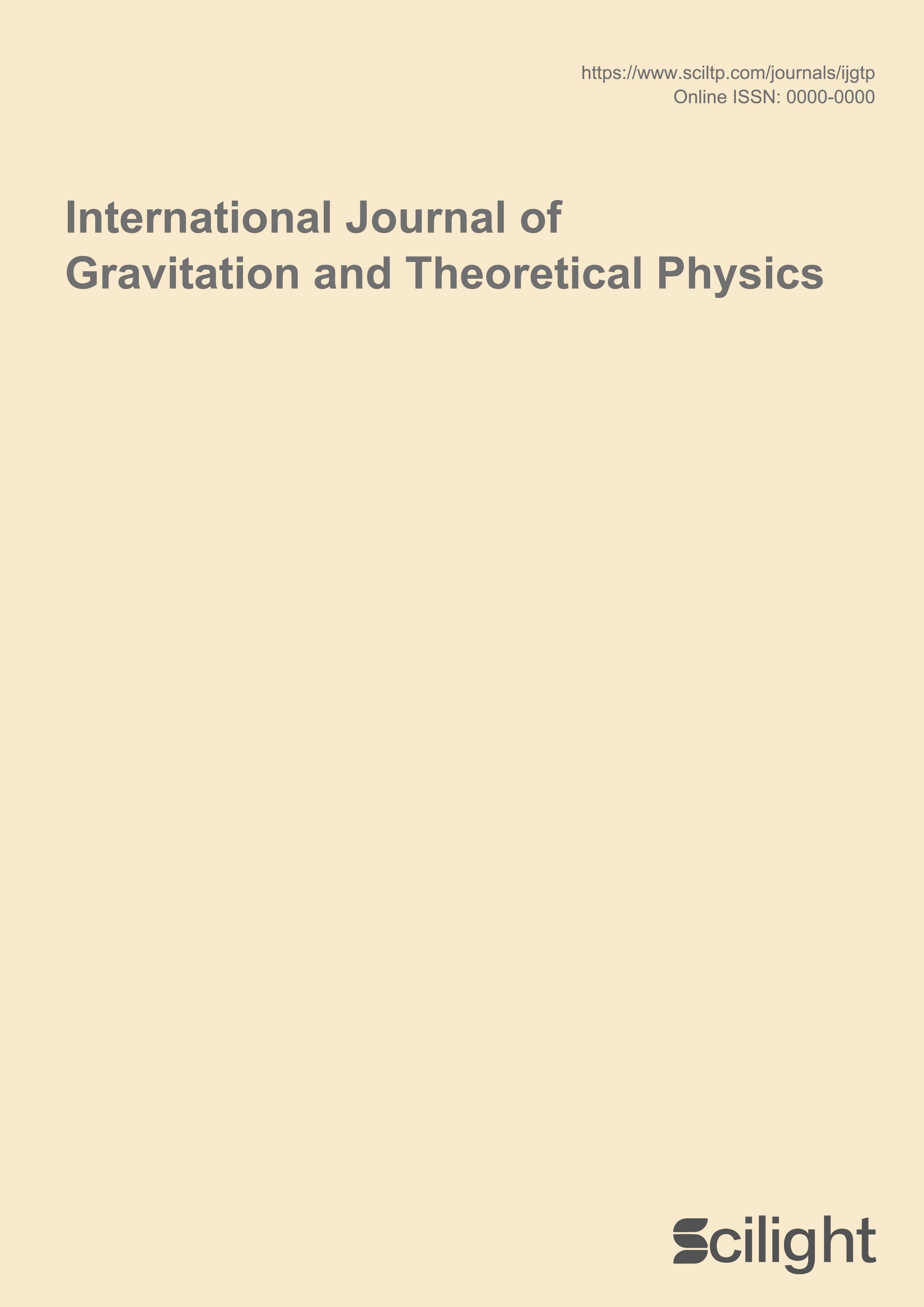} \vspace{-3pt}\\
        \end{tabular}
        \vspace{-22pt}
    \end{table}}
   
    \fancyfoot[C]{
        \vspace{-1.55cm}
        \begin{table}[H]
            \begin{minipage}[c]{0.15\columnwidth}
                \includegraphics[scale=0.5]{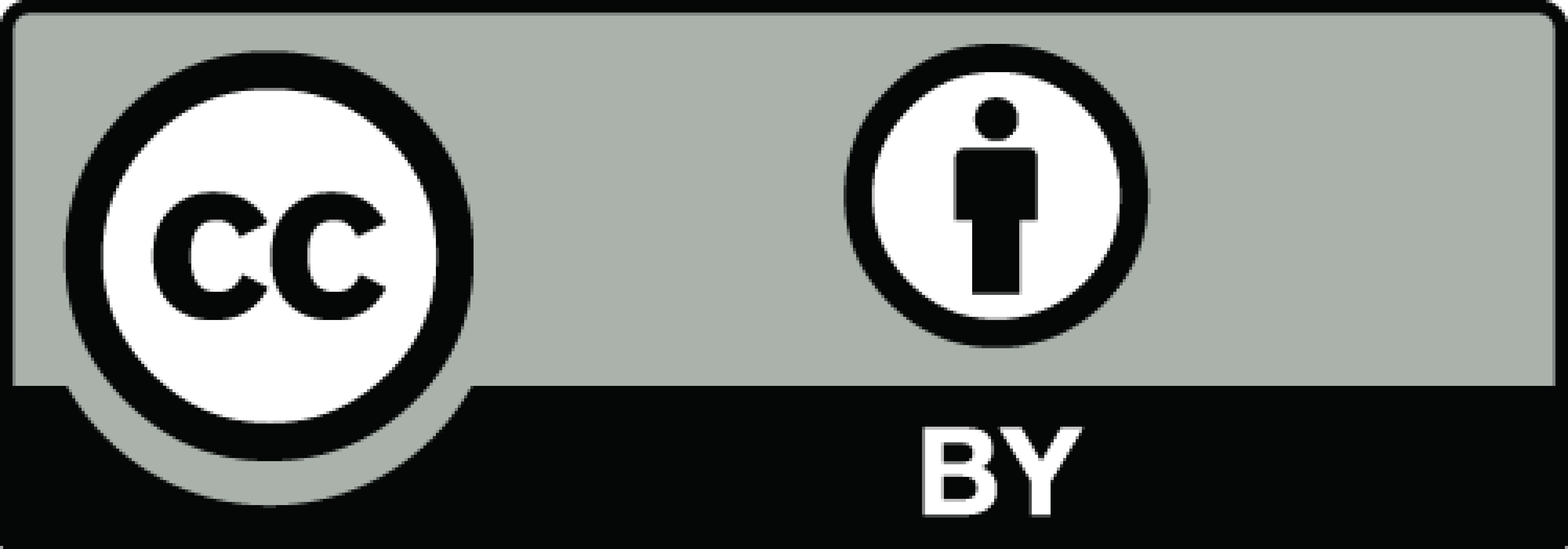} \vspace{1.1pt}
            \end{minipage}
            \hfill
            \begin{minipage}[c]{0.85\columnwidth}
                \scriptsize \textbf{Copyright:} © 2026 by the authors. This is an open access article under the terms and conditions of the Creative Commons Attribution (\mbox{CC BY}) license  (\href{https://creativecommons.org/licenses/by/4.0/}{https://creativecommons.org/licenses/by/4.0/}). \\ \textbf{Publisher’s Note:} Scilight stays neutral with regard to jurisdictional claims in published maps and institutional affiliations.
            \end{minipage}
    \end{table}}
    \vspace{-0.55cm}
}
\setcitestyle{open={[},close={]},citesep={,\!},numbers}
\usepackage[none]{hyphenat}
\begin{document}
\newgeometry{left=2.5cm, right=2.5cm, top=1.8cm, bottom=4cm}
	\thispagestyle{firstpage}
	\nolinenumbers
	{\noindent \textit{Article}}
	\vspace{4pt} \\
	{\fontsize{18pt}{10pt}\textbf{Analytic Expressions for Quasinormal Modes of a Regular Black Hole Sourced by a Dehnen-Type Halo}  }
	\vspace{16pt} \\
	{\large Zainab Malik }
	\vspace{6pt}
	 \begin{spacing}{0.9}
		{\noindent \small
		Institute of Applied Sciences and Intelligent Systems, H-15 Islamabad, Pakistan; zainabmalik8115@outlook.com
			 			\vspace{6pt}\\
		\footnotesize	\textbf{How To Cite}: Malik, Z. Analytic Expressions for Quasinormal Modes of a Regular Black Hole Sourced by a Dehnen-Type Halo. \emph{International Journal of Gravitation and Theoretical Physics} \textbf{2026}, \emph{2}(1), 3. \href{https://doi.org/10.53941/ijgtp.2026.100003}{https://doi.org/10.53941/ijgtp.2026.100003}}\\
	\end{spacing}

\begin{table}[H]
\noindent\rule[0.15\baselineskip]{\textwidth}{0.5pt}
\begin{tabular}{lp{12cm}}
 \small
  \begin{tabular}[t]{@{}l@{}}
  \footnotesize  Received: 15 February 2026 \\
  \footnotesize  Revised: 5 March 2026 \\
   \footnotesize Accepted: 11 March 2026 \\
  \footnotesize  Published: 18 March 2026
  \end{tabular} &
  \textbf{Abstract:} Using an expansion beyond the eikonal regime, we derive relatively compact and accurate analytic expressions for the gravitational quasinormal modes of an asymptotically flat black hole supported by a Dehnen-type dark-matter halo. The spacetime admits a simple analytic metric describing a supermassive black hole embedded in a galactic environment, with the lapse function $f(r)=1-\frac{2 M r^{2}}{(r+a)^{3}}.$ The parameter $a$ sets the characteristic scale of the surrounding halo and controls the regularization of the central region. The axial gravitational sector splits into two distinct channels, referred to as the ``up'' and ``down'' perturbations, which are not isospectral. \\
\\
  &
\end{tabular}
\noindent\rule[0.15\baselineskip]{\textwidth}{0.5pt}
\end{table}

\section{Introduction}

Quasinormal modes (QNMs) describe the characteristic damped oscillations of perturbed black-hole spacetimes and represent one of the most fundamental dynamical signatures of compact objects in general relativity \cite{Kokkotas:1999bd, Konoplya:2011qq, Berti:2009kk, Bolokhov:2025uxz} . They dominate the ringdown phase of gravitational-wave signals and encode detailed information about the background geometry, such as mass, spin, and possible environmental effects \cite{LIGOScientific:2016aoc, LIGOScientific:2017vwq, LIGOScientific:2020zkf}. Since QNM frequencies are determined by the effective potential governing perturbations together with the asymptotic structure of spacetime, they provide a powerful probe of strong-field gravity and potential deviations from vacuum solutions. In parallel, grey-body factors quantify the transmission and reflection of waves through the effective potential barrier and determine scattering properties and the spectrum of Hawking radiation, thereby linking classical perturbation theory with observable astrophysical phenomena.

In realistic astrophysical scenarios, black holes are not isolated systems but are embedded in galactic environments dominated by dark matter. Observational data strongly support the existence of extended dark-matter halos described by empirical density profiles such as Dehnen-type and Einasto distributions, which successfully reproduce galactic rotation curves and large-scale structure properties. While the Schwarzschild and Kerr metrics provide accurate descriptions of isolated black holes, incorporating realistic matter distributions leads to modified geometries that may leave detectable imprints on geodesics, shadows, and wave propagation.

Recently, exact regular black-hole solutions sourced by dark-matter halos have been constructed by adopting the equation of state $P_r = -\rho$ within an anisotropic fluid framework consistent with halo phenomenology \cite{Konoplya:2025ect}. These asymptotically flat solutions are free of curvature singularities and can be obtained analytically for a broad class of galactic density profiles, including Dehnen-type and Einasto models. Although sometimes regularity is achieved via quantum corrections to the theory \cite{Spina:2025wxb}, here the distribution of the environmental matter provides this condition. The resulting metrics provide a physically motivated description of a singularity-free black hole embedded in a galactic halo and offer a natural setting for studying classical and wave dynamics in non-vacuum spacetimes.

Perturbations, stability, quasinormal and scattering of various fields in the vicinity of black holes immersed in astrophysical (galactic) environment have been extensively studied in numerous papers (see, for example,
\cite{Malik:2025czt,Zhang:2021bdr,Pezzella:2024tkf,Konoplya:2021ube,Liu:2024xcd,Daghigh:2022pcr,Tovar:2025apz,Chakraborty:2024gcr,Feng:2025iao,Hamil:2025pte,Mollicone:2024lxy,Zhao:2023tyo,Liu:2024bfj,Dubinsky:2025fwv,Pathrikar:2025sin,Lutfuoglu:2025kqp}).  In the present work, we investigate the quasinormal spectrum of test fields propagating in these halo-sourced geometries. Quasinormal modes of the above black holes have recently been computed in \cite{Bolokhov:2025fto,Lutfuoglu:2025mqa,Saka:2025xxl} using various numerical techniques. This is not surprising, since even the Schwarzschild~solution~does not admit closed-form analytic solutions for the perturbation equations, and in many cases even simpler lower-dimensional black-hole spacetimes \cite{Konoplya:2020ibi} require numerical treatment \cite{Skvortsova:2023zmj,Skvortsova:2023zca}.

Using an analytic expansion in inverse multipole number $1/\ell$, we derive approximate {\it analytic} expressions for QNM frequencies in the high-$\ell$ regime and compare them with precise numerical results. This approach allows us to assess the accuracy of the analytic expansion and to identify qualitative modifications of the spectrum induced by the dark-matter halo. By analyzing how the halo parameters and density profiles affect the quasinormal frequencies and related scattering characteristics, we aim to clarify the role of environmental effects in black-hole spectroscopy.

\section{Black-Hole Background and Master Equations}
\label{sec:wavelike}

In this work we consider a class of static, spherically symmetric, and asymptotically flat black-hole geometries sourced by a Dehnen-type dark-matter halo, recently constructed in~\cite{Konoplya:2025ect}. The solution follows from Einstein gravity coupled to an anisotropic fluid that effectively models a realistic galactic density profile. Remarkably, the metric admits a closed analytic form.

The line element can be written in Schwarzschild-like coordinates as
\begin{equation}
ds^{2} = -f(r)\,dt^{2} + \frac{1}{f(r)}\,dr^{2}
+ r^{2}\left(d\theta^{2}+\sin^{2}\theta\,d\phi^{2}\right),
\label{metric}
\end{equation}
where the lapse function reads
\begin{equation}
f(r) = 1 - 2M r^{2}(r+a)^{-3}.
\label{fmetric}
\end{equation}

Here $M$ denotes the ADM mass of the spacetime, while the positive parameter $a$ sets the characteristic scale of the surrounding halo and encodes the underlying Dehnen density profile~\cite{Dehnen:1993uh,Taylor:2002zd},
\begin{equation}
\rho(r) = \rho_{0}
\left(\frac{r}{a}\right)^{-\alpha}
\left(1+\frac{r^{k}}{a^{k}}\right)^{-(\gamma-\alpha)/k},
\label{Dehnendensity}
\end{equation}
with the parameters fixed to $\gamma=4$, $\alpha=0$, and $k=1$.

The geometry interpolates smoothly between two well-known regimes. At large radii,
\begin{equation}
f(r) = 1 - \frac{2M}{r} + \mathcal{O}\!\left(\frac{1}{r^{2}}\right),
\end{equation}
so that the metric asymptotically approaches Schwarzschild spacetime. In contrast, expanding near the center yields
\begin{equation}
f(r) = 1 - \frac{2M}{a^{3}}\,r^{2} + \mathcal{O}(r^{3}),
\end{equation}
which corresponds to a regular de~Sitter core. Therefore, the solution provides a minimal and physically motivated model of a singularity-free black hole embedded in a galactic halo. Throughout the paper we adopt geometrized units $G=c=1$ and set $M=1$.

%\medskip

\subsection{Axial Perturbations}

The study of linear perturbations in spacetimes supported by anisotropic fluids requires special care. In contrast to vacuum black holes, the background matter sector contributes additional gauge-invariant structures. A systematic treatment of axial (odd-parity) perturbations for anisotropic configurations was presented in~\cite{Chakraborty:2024gcr}, where two consistent perturbative prescriptions were constructed.

The key distinction lies in how perturbations of the fluid four-velocity and stress-energy tensor are constrained. In one formulation, referred to as the ``up'' scheme, the contravariant components of the fluid variables are kept fixed under perturbations. In the alternative ``down'' prescription, the covariant components are held unperturbed instead. Although both approaches are gauge-consistent and physically admissible, they lead to inequivalent master equations because of the intrinsic anisotropy of the fluid.

Expanding the metric perturbations in tensor spherical harmonics and separating the angular dependence through $Y_{\ell m}(\theta,\phi)$, the axial sector reduces, after a Regge--Wheeler-type redefinition of the radial amplitudes, to a Schr\"odinger-like equation,
\begin{equation}
\frac{d^{2}\Psi}{dr_{*}^{2}}
+ \left[\omega^{2}-V(r)\right]\Psi = 0,
\label{eq:master}
\end{equation}
where the tortoise coordinate is defined via
\begin{equation}
r_{*} = \int \frac{dr}{f(r)}.
\end{equation}

The function $\Psi(r)$ represents the gauge-invariant master variable for each multipole $\ell$.

%\medskip

\subsection{Effective Potentials}

The two master equations derived in~\cite{Chakraborty:2024gcr} are characterized by different effective potentials. In terms of the Misner--Sharp mass function $m(r)$ and the fluid variables, they read
\begin{equation}
V^{\text{(up)}}(r)
=
f(r)
\left[
\frac{\ell(\ell+1)}{r^{2}}
-\frac{6\,m(r)}{r^{3}}
+4\pi\bigl(\rho-5P_{r}+4P\bigr)
\right],
\label{Vup_general}
\end{equation}
\begin{equation}
V^{\text{(down)}}(r)
=
f(r)
\left[
\frac{\ell(\ell+1)}{r^{2}}
-\frac{6\,m(r)}{r^{3}}
+4\pi\bigl(\rho-P_{r}\bigr)
\right].
\label{Vdown_general}
\end{equation}

The background stress-energy tensor of the anisotropic halo is diagonal,
\begin{equation}
T^{\mu}{}_{\nu}
=
\mathrm{diag}
\bigl(-8\pi\rho,\; 8\pi P_{r},\; 8\pi P,\; 8\pi P\bigr),
\label{stress-energy}
\end{equation}
where $\rho(r)$, $P_{r}(r)$, and $P(r)$ denote the energy density, radial pressure, and tangential pressure, respectively.

For the specific halo configuration considered in~\cite{Konoplya:2025ect}, the following relations hold:
\begin{equation}
P_{r}(r) = -\rho(r),
\qquad
P(r) = -\rho(r) - \frac{r}{2}\rho'(r).
\label{fluid-conditions}
\end{equation}

Substituting these expressions into Equations~(\ref{Vup_general}) and (\ref{Vdown_general}) yields simplified potentials ~\cite{Konoplya:2025ect},
\begin{eqnarray}
V^{\text{(up)}}(r)
&=&
f(r)
\left[
\frac{\ell(\ell+1)}{r^{2}}
-\frac{6\,m(r)}{r^{3}}
+8\pi\rho(r)
-8\pi r\,\rho'(r)
\right],
\label{Vup_simplified}
\\
V^{\text{(down)}}(r)
&=&
f(r)
\left[
\frac{\ell(\ell+1)}{r^{2}}
-\frac{6\,m(r)}{r^{3}}
+8\pi\rho(r)
\right].
\label{Vdown_simplified}
\end{eqnarray}

In the vacuum limit (when density vanishes), both expressions reduce to the standard Regge--Wheeler potential of Schwarzschild spacetime. However, inside a dark-matter halo the two perturbative prescriptions probe different combinations of the matter variables. This distinction is physically relevant in a realistic astrophysical environment and must therefore be taken into account when computing quasinormal modes, grey-body factors, and absorption cross-sections.
Numerical investigations of quasinormal modes for gravitational perturbations in this background have been reported in \cite{Lutfuoglu:2025mqa,Saka:2025xxl}, whereas quasinormal spectra of test fields were computed in \cite{Bolokhov:2025fto}.

Finally, as demonstrated in~\cite{Konoplya:2025ect}, the effective potentials remain positive definite throughout the exterior region. This guarantees linear stability of axial perturbations and the absence of exponentially growing modes.

\section{The WKB Approach}\label{sec:methodgsforQNMcal}

For the purpose of benchmarking the analytic expressions derived later, we compute the quasinormal frequencies with high numerical accuracy using the WKB approach supplemented by Pad\'e resummation.

Whenever the master equation governing perturbations reduces to a Schr\"odinger-type equation with an effective potential forming a single-peak barrier, the semi-analytic Wentzel--Kramers--Brillouin (WKB) approximation becomes applicable \cite{Iyer:1986np,Konoplya:2003ii,Matyjasek:2017psv}. The method, originally proposed by Schutz and Will and subsequently developed to higher orders, provides approximate complex eigenvalues corresponding to quasinormal oscillations (for recent applications see, e.g., \cite{Bolokhov:2022rqv,Skvortsova:2024atk,Skvortsova:2025cah,Bolokhov:2025lnt,Skvortsova:2024msa,
Dubinsky:2025wns,Pathrikar:2025gzu,Zinhailo:2018ska,Lutfuoglu:2025blw,del-Corral:2022kbk,Bolokhov:2025egl,Skvortsova:2024wly,Lutfuoglu:2025pzi,
Lutfuoglu:2025ljm,Skvortsova:2023zmj,Zhao:2022gxl,Arbelaez:2025gwj,Arbelaez:2026eaz}).

At leading order the WKB approximation reproduces the eikonal (large-$\ell$) limit. To improve the precision for moderate multipole numbers, higher-order corrections are incorporated systematically. In a compact representation, the squared frequency can be expressed as
\begin{equation}
\omega^{2}
=
V_{0}
-
i\left(n+\frac{1}{2}\right)\sqrt{-2 V_{2}}
+
\sum_{p=2}^{2N}
\Lambda_{p},
\label{eq:WKBrewritten}
\end{equation}
where the real and imaginary correction terms are grouped into the coefficients $\Lambda_{p}$. Here
\[
V_{k}
=
\left.
\frac{d^{k}V}{dr_{*}^{k}}
\right|_{r_{*}=r_{*}(r_{0})}
\]
denotes the $k$-th derivative of the effective potential evaluated at its maximum $r_{0}$, and $n=0,1,2,\ldots$ is the overtone number. The correction terms $\Lambda_{p}$ depend on derivatives of the potential up to order $2p$ and encode successive refinements of the approximation.

It is well known that the WKB expansion is asymptotic rather than convergent, and therefore truncating the series at finite order does not guarantee uniform accuracy. A substantial improvement can be achieved by performing a Pad\'e resummation of the truncated expansion. Instead of treating the WKB result as a polynomial in the expansion parameter, one constructs a rational  Pad\'e approximation as prescribed in \cite{Matyjasek:2017psv}.

In practical applications, one compares Pad\'e approximants of neighboring orders and monitors the dispersion among the resulting frequencies to estimate the theoretical uncertainty. Detailed guidelines for assessing reliability and estimating errors can be found in the original references. In the present analysis we employ the sixth-order WKB expansion together with Pad\'e resummation, which provides a good balance between computational efficiency and numerical accuracy.

\section{Expansion beyond the Eikonal Limit}\label{sec:beyondeikonal}

Most publications deriving analytic formulas for quasinormal mode frequencies focus on the eikonal regime for various reasons. The eikonal regime can introduce unusual instabilities in the system \cite{Takahashi:2011qda,Dotti:2004sh,Gleiser:2005ra}. However, its most well-known application is the correspondence between null geodesics and quasinormal modes in this regime \cite{Cardoso:2008bp}. In this work, we begin with a general expansion beyond the eikonal regime and compare our results with numerical data \cite{Lutfuoglu:2025mqa}.

Following \cite{Konoplya:2023moy}, perturbations in a spherically symmetric background can be reduced to the wave-like equation with the effective potential. The latter can be written in the following approximate way:
\begin{equation}\label{potential-multipole}
V(r_*)=\kappa^2\left(H_{0}(r_*)+H_{1}(r_*) \kappa^{-1} +H_{2}(r_*) \kappa^{-2}+\ldots\right).
\end{equation}

Here $\kappa\equiv\ell+\frac{1}{2}$ and $\ell=s,s+1,s+2,\ldots$ is the positive half(integer) multipole number, which has minimal value equal to the spin of the field under consideration $s$. Here, following \cite{Konoplya:2023moy} we use an expansion in terms of $\kappa^{-1}$.

The function $H(r_*)$ possesses a single maximum, and the location of this maximum, denoted by $r_{\max}$, can be expressed as a power series in $\kappa^{-1}$:
\begin{equation}\label{rmax}
  r_{\max} = r_0 + \frac{r_1}{\kappa} + \frac{r_2}{\kappa^2} + \ldots.
\end{equation}

Substituting the series expansion \eqref{rmax} into the first-order WKB approximation:
\begin{eqnarray}
\omega &=& \sqrt{V_0 - i \K \sqrt{-2V_2}},
\end{eqnarray}
and expanding the resulting expression in terms of $\kappa^{-1}$, we obtain:
\begin{eqnarray}\label{eikonal-formulas}
\omega = \Omega \kappa - i \lambda \K + \mathcal{O}(\kappa^{-1}).
\end{eqnarray}

This relation provides an accurate approximation for $\kappa \gg \K$, where $\Omega$ and $\lambda$ represent leading-order contributions to the oscillation frequency and damping rate, respectively. The series expansion underscores how the WKB method captures the dominant behavior of quasinormal modes in the eikonal limit, making it particularly effective in scenarios where the eikonal parameter $\kappa$ is large relative to $\K$. Using this approach the analytic expressions for frequencies have been derived for other black hole models in
\cite{Malik:2024nhy,Malik:2024tuf,Malik:2024sxv,Malik:2024voy,Malik:2023bxc,Malik:2024itg}. Various analytic formulas for quasinormal modes in the eikonal and beyond eikonal limit were reviewed in \cite{Bolokhov:2025uxz}.

%\begin{widetext}
Using the series expansion in terms of the inverse multipole number \cite{Konoplya:2023moy}, for the up potential we find the expansion for the location of the potential peak,
\begin{equation}\label{rmax-up}
\begin{array}{rcl}
r_{\max } &=& \displaystyle\frac{M}{\kappa ^2}+3 M
%\\&&\displaystyle
+a \left(-\frac{14}{3 \kappa ^2}-4\right)
%\\&&\displaystyle
+a^2 \left(\frac{2}{3 M \kappa ^2}-\frac{2}{M}\right)\\
&&\displaystyle+a^3 \left(-\frac{4}{27 M^2 \kappa ^2}-\frac{28}{9 M^2}\right)
%\\&&\displaystyle
+a^4 \left(-\frac{35}{27 M^3 \kappa ^2}-\frac{55}{9 M^3}\right)+\mathcal{O}\left(a^5,\frac{1}{\kappa ^4}\right)
\end{array}
\end{equation}
and, using the WKB formula, the expression
{\footnotesize
\begin{equation}\label{eikonal-up}
\begin{array}{rcl}
\omega  &=& \displaystyle-\frac{i K \left(940 K^2-6599\right)}{46656 \sqrt{3} M \kappa ^2}-\frac{60 K^2+547}{1296 \sqrt{3} M \kappa }+\frac{\kappa }{3 \sqrt{3} M}-\frac{i K}{3 \sqrt{3} M}\\
&&\displaystyle+a \left(\frac{i K \left(1100 K^2-10063\right)}{46656 \sqrt{3} M^2 \kappa ^2}+\frac{204 K^2+319}{3888 \sqrt{3} M^2 \kappa }+\frac{\kappa }{3 \sqrt{3} M^2}-\frac{i K}{9 \sqrt{3} M^2}\right)\\
&&\displaystyle+a^2 \left(\frac{7 i K \left(1852 K^2-13499\right)}{279936 \sqrt{3} M^3 \kappa ^2}+\frac{708 K^2+2941}{7776 \sqrt{3} M^3 \kappa }+\frac{\kappa }{2 \sqrt{3} M^3}+\frac{i K}{54 \sqrt{3} M^3}\right)\\
&&\displaystyle+a^3 \left(\frac{i K \left(214660 K^2-1155893\right)}{2519424 \sqrt{3} M^4 \kappa ^2}+\frac{3684 K^2+22349}{23328 \sqrt{3} M^4 \kappa }+\frac{49 \kappa }{54 \sqrt{3} M^4}+\frac{47 i K}{162 \sqrt{3} M^4}\right)\\
&&\displaystyle+a^4 \left(\frac{i K \left(1755220 K^2-5272121\right)}{10077696 \sqrt{3} M^5 \kappa ^2}+\frac{230700 K^2+1970159}{839808 \sqrt{3} M^5 \kappa }+\frac{1193 \kappa }{648 \sqrt{3} M^5}+\frac{671 i K}{648 \sqrt{3} M^5}\right)+\mathcal{O}\left(a^5,\frac{1}{\kappa ^3}\right)
\end{array}
\end{equation}}
for the quasinormal modes (where $\kappa\equiv\ell+1/2$, $K\equiv n+1/2$).

In a similar way, the position of the maximum of down-potential is
\begin{equation}\label{rmax-down}
\begin{array}{rcl}
r_{\max } &=& \displaystyle\frac{M}{\kappa ^2}+3 M
%\\&&\displaystyle
+a \left(-\frac{2}{\kappa ^2}-4\right)
%\\&&\displaystyle
+a^2 \left(-\frac{2}{3 M \kappa ^2}-\frac{2}{M}\right)\\
&&\displaystyle+a^3 \left(-\frac{28}{27 M^2 \kappa ^2}-\frac{28}{9 M^2}\right)
%\\&&\displaystyle
+a^4 \left(-\frac{55}{27 M^3 \kappa ^2}-\frac{55}{9 M^3}\right)+\mathcal{O}\left(a^5,\frac{1}{\kappa ^4}\right)
\end{array}
\end{equation}
and, using the WKB formula and further expansion in terms of $\kappa$, the expression for quasinormal modes for the down-potential,
{\footnotesize
\begin{equation}\label{eikonal-down}
\begin{array}{rcl}
\omega  &=& \displaystyle-\frac{i K \left(940 K^2-6599\right)}{46656 \sqrt{3} M \kappa ^2}-\frac{60 K^2+547}{1296 \sqrt{3} M \kappa }+\frac{\kappa }{3 \sqrt{3} M}-\frac{i K}{3 \sqrt{3} M}\\
&&\displaystyle+a \left(\frac{i K \left(1100 K^2-3151\right)}{46656 \sqrt{3} M^2 \kappa ^2}+\frac{204 K^2-1409}{3888 \sqrt{3} M^2 \kappa }+\frac{\kappa }{3 \sqrt{3} M^2}-\frac{i K}{9 \sqrt{3} M^2}\right)\\
&&\displaystyle+a^2 \left(\frac{i K \left(12964 K^2-53021\right)}{279936 \sqrt{3} M^3 \kappa ^2}+\frac{708 K^2-3971}{7776 \sqrt{3} M^3 \kappa }+\frac{\kappa }{2 \sqrt{3} M^3}+\frac{i K}{54 \sqrt{3} M^3}\right)\\
&&\displaystyle+a^3 \left(\frac{i K \left(214660 K^2-1010741\right)}{2519424 \sqrt{3} M^4 \kappa ^2}+\frac{3684 K^2-20851}{23328 \sqrt{3} M^4 \kappa }+\frac{49 \kappa }{54 \sqrt{3} M^4}+\frac{47 i K}{162 \sqrt{3} M^4}\right)\\
&&\displaystyle+a^4 \left(\frac{5 i K \left(351044 K^2-1734565\right)}{10077696 \sqrt{3} M^5 \kappa ^2}+\frac{230700 K^2-1485841}{839808 \sqrt{3} M^5 \kappa }+\frac{1193 \kappa }{648 \sqrt{3} M^5}+\frac{671 i K}{648 \sqrt{3} M^5}\right)+\mathcal{O}\left(a^5,\frac{1}{\kappa ^3}\right).
\end{array}
\end{equation}}

In the eikonal limit, the analytic formula derived in~\cite{Bolokhov:2025fto} for test fields
\begin{eqnarray}\nonumber
&&
\omega = \left(\ell+\frac{1}{2}\right) \left(\frac{1}{3 \sqrt{3} M}+\frac{a}{3 \sqrt{3}
   M^2}+\frac{a^2}{2 \sqrt{3}
   M^3}\right)
%   \\\nonumber&&\!\!\!\!\!\!
   -i \left(n+\frac{1}{2}\right) \left(\frac{1}{3
   \sqrt{3} M}+\frac{a}{9 \sqrt{3} M^2}-\frac{a^2}{54 \sqrt{3}
   M^3}\right).
\end{eqnarray}
is accurately reproduced, because in this limit quasinormal modes usually do not depend on the spin of the field. In~the limit $a \rightarrow 0$, the Schwarzschild limit for the metric function is reproduced and the above expression for $\omega$ goes over into the corresponding well-known eikonal expression for quasinormal modes  of the Schwarzschild black hole \cite{Blome:1981azp}.
%\end{widetext}

In the eikonal limit, the derived formulas formally correspond to the WKB frequencies that describe late-time decay. However, as demonstrated in \cite{Konoplya:2017wot,Bolokhov:2023dxq,Khanna:2016yow}, the WKB approach may be unable to describe the quasinormal spectrum even in the regime of high multipole numbers. Consequently, the relationship connecting the angular velocity $\Omega$ at unstable null geodesics and the Lyapunov exponent $\lambda$ with the eikonal quasinormal frequencies \cite{Cardoso:2008bp}:
\begin{equation}\label{QNM}
\omega_n = \Omega\left(\ell+\tfrac{1}{2}\right) - i\left(n+\tfrac{1}{2}\right)|\lambda|, \quad \ell \gg n,
\end{equation}
holds true, but only for the spectrum branch (if any) that the WKB method can reproduce. Here we can see that the correspondence holds for both types of gravitational perturbations.

The quasinormal spectrum exhibits a clear and systematic dependence on the halo parameter $a$. As $a$ increases, the real part of the frequency grows monotonically for both the ``up'' and ``down'' gravitational channels, indicating that the presence of the halo effectively stiffens the potential barrier and raises the oscillation frequency. At the same time, the imaginary part shows a more moderate variation, typically becoming slightly more negative in magnitude, which corresponds to a mild increase in the damping rate. Thus, the primary effect of the halo parameter is a shift in the oscillation frequency, while the damping rate remains comparatively less sensitive.

A comparison between the analytic expressions obtained via the $1/\ell$ expansion and the numerical WKB results demonstrates very good agreement. For $\ell=3$ {(Table~\ref{tabl:l3})}, the relative deviation remains below the percent level over the considered range of $a$, confirming that the analytic approximation provides highly accurate predictions already at moderately large multipole numbers. Even for $\ell=2$ {(Table~\ref{tabl:l2})}, where the expansion is formally less justified, the discrepancy remains at the level of a few percent. Importantly, this deviation is smaller than the overall variation of the frequencies caused by changing $a$, so the analytic approximation reliably captures the halo-induced modification of the spectrum.
\vspace{-12pt}
\begin{table}[H]
\centering
\caption{Quasinormal modes of the $\ell=3$ ($M=1$) calculated using the WKB formula at the sixth order and analytic~approximation.}\label{tabl:l3}
\footnotesize
\scalebox{.88}[1.0]{\begin{tabular}{>{\centering\arraybackslash}m{0.03\textwidth}		
                >{\centering\arraybackslash}m{0.19\textwidth}		
                >{\centering\arraybackslash}m{0.19\textwidth}
                >{\centering\arraybackslash}m{0.08\textwidth}		
                >{\centering\arraybackslash}m{0.19\textwidth}		
                >{\centering\arraybackslash}m{0.19\textwidth}
                >{\centering\arraybackslash}m{0.08\textwidth}}	
\toprule	
\multicolumn{4}{c}{\textbf{``up''}}&\multicolumn{3}{c}{\textbf{``down''}}\\
\midrule
\boldmath{$a$} & \textbf{WKB-6 (\boldmath{$m=3$})} & \textbf{Analytic} & \textbf{Difference} & \textbf{WKB-6 (\boldmath{$m=3$})} & \textbf{Analytic} & \textbf{Difference} \\
\hline
$0$     & $0.5994434-0.0927029 i$ & $0.6020429-0.0930107 i$ & $0.432\%$ & $0.5994434-0.0927029 i$ & $0.6020429-0.0930107 i$ & $0.432\%$\\
$0.02$  & $0.6136579-0.0934210 i$ & $0.6162757-0.0937514 i$ & $0.425\%$ & $0.6120647-0.0933597 i$ & $0.6147482-0.0936801 i$ & $0.436\%$\\
$0.04$  & $0.6288476-0.0941408 i$ & $0.6314747-0.0944897 i$ & $0.417\%$ & $0.6255137-0.0940110 i$ & $0.6282862-0.0943444 i$ & $0.441\%$\\
$0.06$  & $0.6451385-0.0948571 i$ & $0.6477600-0.0952206 i$ & $0.406\%$ & $0.6398952-0.0946512 i$ & $0.6427580-0.0949984 i$ & $0.446\%$\\
$0.08$  & $0.6626828-0.0955626 i$ & $0.6652662-0.0959377 i$ & $0.390\%$ & $0.6553349-0.0952724 i$ & $0.6582784-0.0956357 i$ & $0.448\%$\\
$0.1$   & $0.6816663-0.0962467 i$ & $0.6841454-0.0966337 i$ & $0.364\%$ & $0.6719861-0.0958639 i$ & $0.6749743-0.0962491 i$ & $0.444\%$\\
$0.12$  & $0.7023196-0.0968945 i$ & $0.7045640-0.0973000 i$ & $0.322\%$ & $0.6900380-0.0964107 i$ & $0.6929870-0.0968301 i$ & $0.427\%$\\
$0.14$  & $0.7249334-0.0974842 i$ & $0.7267059-0.0979271 i$ & $0.250\%$ & $0.7097286-0.0968912 i$ & $0.7124690-0.0973692 i$ & $0.388\%$\\
$0.16$  & $0.7498828-0.0979833 i$ & $0.7507689-0.0985042 i$ & $0.136\%$ & $0.7313634-0.0972739 i$ & $0.7335888-0.0978559 i$ & $0.312\%$\\
$0.18$  & $0.7776644-0.0983420 i$ & $0.7769700-0.0990190 i$ & $0.124\%$ & $0.7553447-0.0975109 i$ & $0.7565300-0.0982780 i$ & $0.185\%$\\
$0.2$   & $0.8089570-0.0984807 i$ & $0.8055330-0.0994601 i$ & $0.437\%$ & $0.7822185-0.0975271 i$ & $0.7814750-0.0986241 i$ & $0.168\%$\\
$0.22$  & $0.8447280-0.0982652 i$ & $0.8367105-0.0998121 i$ & $0.960\%$ & $0.8127574-0.0971974 i$ & $0.8086416-0.0988792 i$ & $0.543\%$\\
$0.24$  & $0.8864359-0.0974514 i$ & $0.8707620-0.1000602 i$ & $1.78\%$  & $0.8481131-0.0962975 i$ & $0.8382460-0.0990289 i$ & $1.20\%$\\
%$0.26$ & $0.9179902-0.0936226 i$ & $0.8891468-0.0988161 i$ & $3.18\%$  & $0.8901372-0.0943783 i$ & $0.8705216-0.0990571 i$ & $2.25\%$\\
%$0.28$ & $0.9804855-0.0888796 i$ & $0.9276594-0.0986410 i$ & $5.46\%$  & $0.9421602-0.0903623 i$ & $0.9057145-0.0989468 i$ & $3.96\%$\\
\bottomrule
\end{tabular}}
\end{table}
\vspace{-18pt}
\begin{table}[H]
\centering
\caption{Quasinormal modes of the $\ell=2$ ($M=1$)  calculated using the WKB formula at the sixth order and analytic~approximation.}\label{tabl:l2}
%\begin{tabular}{l | c c l | c c l}
\footnotesize
\scalebox{.88}[1.0]{\begin{tabular}{>{\centering\arraybackslash}m{0.03\textwidth}		
                >{\centering\arraybackslash}m{0.19\textwidth}		
                >{\centering\arraybackslash}m{0.19\textwidth}
                >{\centering\arraybackslash}m{0.08\textwidth}		
                >{\centering\arraybackslash}m{0.19\textwidth}		
                >{\centering\arraybackslash}m{0.19\textwidth}
                >{\centering\arraybackslash}m{0.08\textwidth}}	
\toprule	
\multicolumn{4}{c}{\textbf{``up''}}&\multicolumn{3}{c}{\textbf{``down''}}\\
\midrule
\boldmath{$a$} & \textbf{WKB-6 (\boldmath{$m=3$})} & \textbf{Analytic} & \textbf{Difference} & \textbf{WKB-6 (\boldmath{$m=3$})} & \textbf{Analytic} & \textbf{Difference} \\
\hline
$0$     & $0.3736199-0.0889328 i$ & $0.3809799-0.0899249 i$ & $1.93\%$  & $0.3736199-0.0889328 i$ & $0.3809799-0.0899249 i$ & $1.93\%$\\
$0.02$  & $0.3839877-0.0896214 i$ & $0.3913805-0.0907635 i$ & $1.90\%$  & $0.3816645-0.0895694 i$ & $0.3892420-0.0906239 i$ & $1.95\%$\\
$0.04$  & $0.3950753-0.0903329 i$ & $0.4025137-0.0916062 i$ & $1.86\%$  & $0.3902288-0.0902091 i$ & $0.3980498-0.0913214 i$ & $1.97\%$\\
$0.06$  & $0.4070179-0.0910701 i$ & $0.4144720-0.0924484 i$ & $1.82\%$  & $0.3993786-0.0908392 i$ & $0.4074700-0.0920128 i$ & $2.00\%$\\
$0.08$  & $0.4199295-0.0918148 i$ & $0.4273598-0.0932843 i$ & $1.76\%$  & $0.4091983-0.0914469 i$ & $0.4175769-0.0926924 i$ & $2.02\%$\\
$0.1$   & $0.4339700-0.0925714 i$ & $0.4412946-0.0941072 i$ & $1.69\%$  & $0.4197974-0.0920216 i$ & $0.4284551-0.0933534 i$ & $2.04\%$\\
$0.12$  & $0.4493251-0.0933313 i$ & $0.4564060-0.0949093 i$ & $1.58\%$  & $0.4312904-0.0925722 i$ & $0.4401970-0.0939883 i$ & $2.04\%$\\
$0.14$  & $0.4662235-0.0940611 i$ & $0.4728354-0.0956816 i$ & $1.43\%$  & $0.4438166-0.0930546 i$ & $0.4529037-0.0945882 i$ & $2.03\%$\\
$0.16$  & $0.4849706-0.0947552 i$ & $0.4907373-0.0964142 i$ & $1.21\%$  & $0.4575866-0.0934419 i$ & $0.4666852-0.0951435 i$ & $1.98\%$\\
$0.18$  & $0.5059815-0.0953532 i$ & $0.5102780-0.0970960 i$ & $0.901\%$ & $0.4728569-0.0936933 i$ & $0.4816600-0.0956430 i$ & $1.87\%$\\
$0.2$   & $0.5297927-0.0957949 i$ & $0.5316367-0.0977151 i$ & $0.494\%$ & $0.4899763-0.0937387 i$ & $0.4979556-0.0960765 i$ & $1.67\%$\\
$0.22$  & $0.5572195-0.0959174 i$ & $0.5550043-0.0982581 i$ & $0.570\%$ & $0.5094378-0.0934597 i$ & $0.5157078-0.0964298 i$ & $1.34\%$\\
$0.24$  & $0.5894736-0.0954847 i$ & $0.5805840-0.0987111 i$ & $1.58\%$  & $0.5319761-0.0926420 i$ & $0.5350610-0.0966896 i$ & $0.943\%$\\
%$0.26$ & $0.6015909-0.0897065 i$ & $0.5822451-0.0963691 i$ & $3.36\%$  & $0.5587646-0.0908387 i$ & $0.5561699-0.0968413 i$ & $1.16\%$\\
%$0.28$ & $0.6518269-0.0851445 i$ & $0.6099181-0.0962697 i$ & $6.60\%$  & $0.5918319-0.0870297 i$ & $0.5791952-0.0968692 i$ & $2.68\%$\\
\bottomrule
\end{tabular}}
\end{table}

\section{Conclusions}

In this work, we derived analytic expressions for the gravitational quasinormal modes of an asymptotically flat regular black hole supported by a Dehnen-type dark-matter halo, using an expansion in inverse multipole number $1/\ell$. The obtained formulas provide compact and transparent representations of the spectrum and allow for a direct assessment of the influence of the halo parameter $a$.

Our analysis shows that the presence of the halo systematically modifies the quasinormal frequencies. In particular, increasing $a$ leads to a monotonic growth of the real part of the modes, while the imaginary part exhibits a comparatively weaker dependence. Thus, the dominant halo effect is a shift of the oscillation frequency, with only moderate changes in the damping rate.

A detailed comparison with numerical WKB results demonstrates that the analytic approximation is highly accurate for $\ell \geq 3$, where the relative deviation remains at the percent level or below. Even for $\ell=2$, the error is small compared to the overall variation of the frequencies induced by the halo parameter. Therefore, the $1/\ell$ expansion provides a reliable and efficient tool for describing gravitational quasinormal modes in black holes embedded in realistic galactic environments.

%		\section*{Institutional Review Board Statement}
%Not applicable.

%		\section*{Informed Consent Statement}
%Not applicable.

%		\section*{Data Availability Statement}
%Not applicable.

\section*{Acknowledgments}
The author would like to acknowledge Alexander Zhidenko for useful comments.

		\section*{Conflicts of Interest}

The author declares no conflict of interest.

		\section*{Use of AI and AI-Assisted Technologies}

During the preparation of this work, the author used chatGPT to refine language. After using this service, the author reviewed and edited the content as needed and takes full responsibility for the content of the published article.

\end{document}